\documentclass[12pt,preprint]{aastex}

\def\casp{Comments.Astrophys.Space Phys. \,}

\def\physrep{Phys.Rep. \,}

\def\bs{\bigskip}

\def\ea{et al. \,}
\def\eg{{\it e.g.\,}}
\def\be{\begin{equation}}
\def\ee{\end{equation}}
\def\rel{relativistic \,}

\def\sz{Sunyaev \& Zeldovich \,}
 
\begin{document}
\bs
\bs

\title{CMB Comptonization by Energetic Nonthermal Electrons \\in 
Clusters of Galaxies}
\vspace{2cm}
\bs

\author{\bf Meir Shimon}
\affil{School of Physics and Astronomy, Tel Aviv University, Tel 
Aviv, 69978, Israel}

\email{meirs@ccsg.tau.ac.il}

\vspace{2cm}

\and

\vspace{2cm}

\author{\bf Yoel Rephaeli}

\affil{School of Physics and Astronomy, Tel Aviv University, Tel 
Aviv, 69978, Israel, \\and\\ Center for Astrophysics and Space 
Sciences, University of California, San Diego, La Jolla, 
CA\,92093-0424}

\email{yoelr@noga.tau.ac.il}

\begin{abstract}

Use of the Sunyaev-Zeldovich effect as a precise cosmological probe
necessitates a realistic assessment of all possible contributions to
Comptonization of the cosmic microwave background in clusters of 
galaxies. We have calculated the additional intensity change due to 
various possible populations of energetic electrons that have been 
proposed in order to account for measurements of intracluster radio,
nonthermal X-ray and (possibly also) EUV emission. Our properly 
normalized estimates of (the highly model dependent value of) the 
predicted intensity change due to these electrons is well below 
$\sim 6\%$ and $\sim 35\%$ of the usual Sunyaev-Zeldovich effect due 
to electrons in the hot gas in Coma and A2199, respectively. These 
levels constitute high upper limits since they are based on energetic 
electron populations whose energy densities are {\it comparable} to 
those of the thermal gas. The main impact of nonthermal 
Comptonization is a shift of the crossover frequency (where the thermal 
effect vanishes) to higher values. Such a shift would have important 
consequences for our ability to measure cluster peculiar velocities 
from the kinematic component of the Sunyaev-Zeldovich effect.
\end{abstract}

\section{Introduction}

The Sunyaev-Zeldovich (S-Z) effect is a small intensity change that 
results from Compton scattering of the cosmic microwave background 
(CMB) radiation by electrons in the hot gas in clusters of galaxies 
(Zeldovich \& Sunyaev 1969, \sz 1972). The effect constitutes a unique 
cluster and cosmological probe (for reviews, see Rephaeli 1995a, 
Birkinshaw 1999), whose great potential is beginning to be realized in 
recent years following the major observational progress in obtaining 
sensitive images of the effect by (mostly) interferometric arrays 
(Jones et al. 1993, Carlstrom et al. 2000, Udomprasert, Mason, \&
Readhead 2000, and the review by Carlstrom \ea 2001)
and the increasingly more precise values of the Hubble constant 
($H_0$) that have been deduced from S-Z and X-ray measurements. For 
example, a fit to data from 33 cluster distances 
yields $H_0 = 58$ km s$^{-1}$ Mpc$^{-1}$ (in a flat cosmological 
model), with direct observational errors of $\pm 5\%$ (Carlstrom \ea 
2001), but with a much larger level of systematic uncertainties. Among 
others, the latter include simplified modeling of the properties of 
the hot intracluster (IC) gas, and cluster morphology. These are the 
main limitations to the use of the S-Z effect as a precise cosmological 
probe, and as such give further motivation for in-depth studies of 
the cluster environment. 

Energetic non-thermal (NT) electrons whose pressure is not negligible 
compared to the thermal gas pressure constitute an aspect of IC 
phenomena with possibly appreciable ramifications for precision S-Z 
measurements. The presence of significant energetic electron 
populations in many clusters has been known from measurements of 
diffuse IC radio emission, and recently also from NT X-ray emission 
in a few clusters (Rephaeli, Gruber \& Blanco 1999, hereafter RGB, 
Fusco-Femiano \ea 1999, Kaastra et al. 1999, Gruber \& Rephaeli 2002). NT 
electrons produce an additional degree of Comptonization which 
amounts to a small intensity change ($\Delta I_{nt}$) that must be 
accounted for, particularly in the measurement of $H_0$ from the thermal 
component, and peculiar cluster velocities from the kinematic component 
of the S-Z effect. Relativistic generalizations (Rephaeli 1995b, Challinor 
\& Lasenby 1998, Sazonov \& Sunyaev 1998) of the original non-relativistic 
calculations of \sz (1972, 1980) have now been performed to a sufficiently 
high level of accuracy, including also terms of order $\tau^2$
(Nozawa, Itoh, \& Kohyama 1998, Itoh \ea 2000, Shimon \& Rephaeli
2002), where $\tau$ is the Thomson optical depth of the cluster.
In calculating the effect of multiple scatterings the finite size of the 
cluster has to be explicitly accounted for; this has been done in the 
Monte-Carlo simulations of Molnar \& Birkinshaw (1999).
The \rel treatment provides the theoretical basis for calculation also 
of $\Delta I_{nt}$.

CMB Comptonization by NT electrons was first assessed for conditions 
in lobes (McKinnon, Owen, \& Eilek 1991) and cocoons (Yamada \& Sugiyama 1999)
of radio galaxies, with its possible use to measure their electron 
pressure. The effect of such electrons in clusters was recently 
considered in some more detail (Blasi \& Colafrancesco 1999, Ensslin 
\& Kaiser 2000, Blasi, Olinto, \& Stebbins 2000).
Clearly, the higher the electron
pressure, the higher is the degree of Comptonization induced by the 
electrons, and because models of NT electrons vary greatly in energy 
density, estimates of their impact on the CMB range from a negligible 
level, to a very substantial fraction of the thermal S-Z effect due to 
the hot gas. It is quite essential to study NT electron 
populations in order to {\it realistically} determine the spectral 
and spatial profiles of $\Delta I_{nt}$ in clusters in which extended 
radio emission has been measured. This may lead to ways by which their 
impact on S-Z work can be minimized, and eventually the Comptonization 
induced by NT electrons could perhaps even be used as a diagnostic tool 
of these electrons. 

In this paper we perform an exact calculation of the degree of 
Comptonization predicted in a range of NT electron models that have 
been proposed to explain radio, EUV, and hard X-ray emission in 
clusters. Our estimates of $\Delta I_{nt}$ in Coma and A2199 
are based on recent results on the electron populations in 
these clusters, and the main features of the resulting spectral change 
are contrasted with those of the thermal component of the S-Z effect.

\section{Non-thermal Electron Populations}

The main evidence for \rel electrons and magnetic fields in clusters 
comes from measurements of extended IC regions of radio emission (in 
the frequency range $\sim 0.04-4$ GHz) with spectral indices and 
luminosities in the range $\sim 1-2$, and $10^{40.5}$ -- $10^{42}$ 
erg/s ($H_0$ = 50 km s$^{-1}$ Mpc$^{-1}$), respectively. Many of 
the radio emitting regions detected in over 30 clusters (Giovannini,
Tordi, \& Feretti 1999, Giovannini \& Feretti 2000) are central, with
sizes in the range $\sim 1-3$ Mpc. 
The energy range of the emitting electrons depends on the value of the 
mean, volume-averaged magnetic field, a quantity which is not known 
very well, but is likely to be in the range $\sim 0.1 - 1$ $\mu$G (for 
a recent review, see Rephaeli 2001). The range of electron energies 
implied from these measurements is roughly $1-100$ GeV, but electrons 
with energies both below and above this range are also expected on 
theoretical grounds.

Relativistic electrons produce X-ray emission in a wide spectral 
range by Compton scattering off the CMB (Rephaeli 1979) and by NT 
bremsstrahlung. This emission has quite possibly been measured already 
in Coma (Rephaeli, Gruber \& Blanco 1999, Fusco-Femiano \ea 1999), 
A2199 (Kaastra \ea 1999), A2256 (Molendi, De Grandi, \& Fusco-Femiano
2000), and A2319 (Gruber \& Rephaeli 2002) by the RXTE and BeppoSAX
satellites. These were spectral measurements; 
the PCA, HEXTE (both aboard RXTE) and PDS (aboard BeppoSAX) experiments 
do not have the adequate spatial resolution. Thus, we have no spatial 
information on NT electrons from X-ray measurements, and only rudimentary 
knowledge of the morphology of the radio emission (which, however, involves 
also the unknown spatial distribution of the magnetic field). Therefore, 
in order to avoid the need for introducing unknown parameters, we 
characterize electron populations in terms of the spectral distribution 
of their total number, ignoring their spatial profiles in the central 
$\sim 1$ Mpc region where they mostly reside.   

Of the various proposed NT IC electron models (see, \eg, Sarazin 1999) 
we focus here on those that have been contrasted with actual observational 
data, with the electron distribution appropriately normalized by the 
determination of both their total number and the power-law index, $q$. 
Some of the proposed electron models were motivated by the presumption 
that NT X-ray emission could also originate from NT 
bremsstrahlung by energetic -- though not necessarily highly relativistic 
-- electrons, and by an attempt to account also for observed EUV emission 
from a few clusters. This emission is said to be NT (\eg, Sarazin \& Lieu 
1998, Bowyer \ea 1999), possibly by a population of low energy electrons. 
The full electron distribution could be described either as a sum of two
separate parts (thermal plus NT), or by a `super Maxwellian', consisting
of a truncated Maxwellian with a power law tail (Blasi, Olinto, \&
Stebbins 2000) `sawn' 
together at a given energy (\eg, $\sim 3kT_e$, where $T_e$ is the electron 
[and gas] temperature). We consider here four specific models whose 
parameters in Coma and A2199 have been determined from radio, EUV, and 
NT X-ray measurements. 

The simplest model for the electron momentum distribution is a power-law 
over a sufficiently wide range so as to explain both the observed radio 
and possibly Compton-produced X-ray emission. We express the electron 
spectrum in terms of the (differential) number, $N(p)$, per unit 
dimensionless momentum, $p=\beta\gamma$, where $\gamma$ and $\beta$ are 
the Lorentz factor and dimensionless velocity, $\beta=v/c$,
respectively,
\begin{eqnarray}
N(p)&=&A p^{-q},    \quad   p_1 \leq p  \leq p_2 \cr
\cr
A&=&{N_{12}(q-1) \over p_1 ^{-(q-1)} - p_2 ^{-(q-1)}}.
\end{eqnarray}
The limiting momenta $p_1$ and $p_2$ are the lower and higher values of 
$p$ that correspond to the {\it observed radio frequency range}, and 
$N_{12}$ is the total number of electrons with energies in this 
specific interval, $[p_{1}, p_{2}]$. Details of such a model were 
worked out long ago (Rephaeli 1977, 1979) and will not be repeated here. 
Suffice it to say that the {\it full} momentum range is substantially 
uncertain, especially its low end which is of particular relevance to 
our discussion here. 

A simple way to obtain the lower energy extension of the above 
distribution is to assume that steady state is attained, whereby 
electrons are continually accelerated to compensate for radiative 
energy losses -- defined here in terms of $dp/dt$ -- and a population 
of lower energy electrons is built up. The latter can be determined by 
taking into account the dominant rate of energy loss at energies well 
below $\sim 150$ MeV, electronic (Coulomb) excitations, given by 
(Rephaeli 1979)
\begin{eqnarray}
b_{ee}=1.2\times
10^{-12}n_{e}\left[1.0+\frac{\ln\left(\sqrt{1+p^{2}}/n_{e}\right)}{75}
\right]\rm{s}^{-1},
\end{eqnarray}
where $n_{e}$ is the (thermal) electron number density. At higher 
energies Compton-synchrotron losses dominate; these occur at a combined 
rate (\eg, Rephaeli 1979)
\be
b_{cs}=1.37\times 10^{-20}[(1+z)^{4} + 9.5\times 10^{-2} (B/1\mu\rm{G})^{2}] 
\,  \rm{s}^{-1},
\ee
where $z$ is the cluster redshift, and $B$ is the mean, volume-averaged
value of the magnetic field. The steady state solution is then (Rephaeli 
1979) 
\begin{eqnarray}
N_{I} \left(p\right)&=&\frac{Ab_{cs}}{b\left(p\right)}
\left(1+p^{2}\right)^{-\frac{\left(q-2\right)}{2}}\qquad\left(p_{1}\leq
  p\leq
  p_{2}\right)\nonumber\\&=&\frac{Ab_{cs}}{b\left(p\right)}
\left(1+p_{1}^{2}\right)^{-\frac{\left(q-2\right)}{2}}\qquad\left(p\leq
p_{1}\right),\nonumber\\
\end{eqnarray}
where
$b \equiv dp/dt = b_{ee}+b_{cs}\left(1+p^{2}\right)$.
Clearly, the observed radio frequency range implies that $p_{1}\gg 1$.
(Note that another possible energy loss mechanism is scattering by Alfven 
waves, which the electrons themselves can excite if their spatial 
distribution is somewhat anisotropic. Because of the substantial uncertainty 
in estimating this loss mechanism [\eg, Rephaeli 1979] it will not be 
considered here.) Parameters of this model were determined from RXTE 
measurements of the Coma cluster, assuming the emission is from Compton 
scattering of \rel electrons off the CMB (Rephaeli, Gruber, \& Blanco 1999), 
and from BeppoSAX measurements of A2199 (Kaastra \ea 1999). This 
appropriately extended electron population which has a power-law form 
at low and high energies, but with indices whose values differ by unity, 
is our first model. The parameters of all the models considered here are 
given in Table 1. 

\begin{table*}[h]
\small
\begin{center}
\begin{tabular}{llllll}
\tableline
\tableline
Cluster&Model &Model
details&$\qquad N_{0}$&$\quad N_{nt}^{tot}$&$N_{nt}^{tot}/N_{th}^{tot}$\cr
\tableline
Coma & && &\cr
     &I&Extended power law &$2.275\times 10^{72}$&$1.09\times
     10^{65}$&$4.70\times 10^{-7}$\cr
     &II&Shock accelerated &$4.270\times
     10^{67}$&$2.98\times 10^{68}$&$1.28\times 10^{-3}$\cr
     &III&Cooling electrons &$5.238\times 10^{62}$&$3.13\times
     10^{65}$&$1.35\times 10^{-6}$\cr
     &IV&Simple power law q=3.68&$8.970\times 10^{68}$&$9.27\times
     10^{69}$&$0.04$\cr
&&\cr
A2199 & && &\cr
     &I&Extended power law&$2.306\times 10^{68}$&$1.66\times
     10^{64}$&$2.10\times 10^{-7}$\cr
     &II&Shock accelerated&$4.240\times
     10^{67}$&$6.22\times 10^{68}$&$7.90\times 10^{-3}$\cr
     &III&Cooling electrons &$5.238\times 10^{62}$&$3.13\times
     10^{65}$&$3.95\times 10^{-6}$\cr
     &IV&Power law q=3.33&$5.420\times 10^{68}$&$6.47\times
     10^{69}$&$0.082$\cr
\tableline
\end{tabular}
\end{center}
\caption{Models for NT electrons in the Coma and A2199 clusters.}
\end{table*}

Two other forms for $N(p)$ are discussed by Sarazin \& Kempner (2000). In 
the first model electrons are accelerated to relativistic energies by 
shell-type supernova remnants (Baring \ea 1999). When the back reaction of 
the accelerating electrons on the shock structure is taken into account, 
then this presumably leads to a modified power-law form at small momenta, 
resulting in the following distribution, 
\begin{eqnarray}
N_{II} \left(p\right)=\frac{N_{0}p^{-2}}{1+p_{c}^{2}}\left[1+
\left(\frac{p_{c}}{p}\right)^{2}\right].
\end{eqnarray}
We assume $p_{c}=1$ in normalizing the model.
Explicit account of Coulomb losses yields the third distribution 
\begin{eqnarray}
N_{III}\left(p\right)=\frac{2N_{0}p^{2}}{1+p^{2}}.
\end{eqnarray}
$N_{0}$ is derived from comparison of the models with observations 
and assuming that the hard X-ray emission mechanism is NT bremsstrahlung. 
The electron distribution in Equation (6) has an upper momentum cutoff 
at $p=300$. 

Sarazin \& Kempner (2000) also propose a simple power law model which 
extends to lower energies {\it with no change in the index}. This is the 
fourth (IV) model explored here. The observationally deduced parameters for 
the above four models are listed in Table 1; in these models, the lower 
momentum cutoff is $p_{1}=\sqrt{(1+3\Theta)^{2}-1}$, where 
$\Theta=kT_{e}/mc^{2}$.

\section{Comptonization}

A detailed description of the Comptonized spectrum resulting from 
scattering of the CMB by a thermal, non-\rel population of electrons 
was given by \sz (1972, 1980). A more accurate treatment of the process 
requires \rel generalization due to the rapid motion of electrons in 
the hot ($kT_{e} \leq 15$ keV) IC gas, as has been shown explicitly by 
Rephaeli (1995a). 
A fully relativistic treatment (but still in the Thomson limit, and 
with electron recoil and induced scattering safely ignored)
is obviously required in order to calculate the Comptonized spectrum 
resulting from scattering of the radiation by a NT energetic electron 
population. We use the approach adopted in the latter paper (see also 
Rephaeli \& Yankovitch 1997).

The probability of scattering of an incoming photon moving originally  
in the direction specified by $\mu_{0}=\cos\theta_{0}$, to the 
direction $\mu'_{0}=\cos\theta'_{0}$, is (Chandrasekhar 1950)
\begin{eqnarray}
f\left(\mu_{0},\mu'_{0}\right)=\frac{3}{8}\left[1+
\mu_{0}^{2}\mu'^{2}_{0}+\frac{1}{2}\left(1-\mu_{0}^{2}\right)\left(1-
\mu'^{2}_{0}\right)\right],
\end{eqnarray}
where the subscript $0$ denotes quantities in the electron rest
frame.
The logarithmic frequency shift in the scattering is 
\begin{eqnarray}
s\equiv\ln\left(\nu'/\nu\right)=\ln\left(\frac{1+\beta\mu'_{0}}
{1+\beta\mu_{0}}\right) \,,
\end{eqnarray}
where $\beta$ is the dimensionless electron velocity in the CMB frame. 
The probability for scattering is conveniently written in 
terms of the variables $\beta$ and $t=e^{s}$ (Wright 1979), 
\begin{eqnarray}
\mathcal{P} \left(s,\beta\right)=\frac{1}{2\gamma^{4}\beta}
\int_{\mu_{1}}^{\mu_{2}}\frac{e^{s}f\left(\mu_{0},
\mu'_{0}\right)}{\left(1+\beta\mu_{0}\right)^{2}}d\mu_{0},
\end{eqnarray}
where $\mu_{1}$ and $\mu_{2}$ are given by
\begin{eqnarray}
\mu_{1}=\left\{
\matrix{\frac{e^{-s}\left(1-\beta\right)-1}{\beta}&s\leq 0
\cr
   -1&s\geq 0}\right.
\end{eqnarray}
\begin{eqnarray}
\mu_{2}=\left\{
\matrix{1&s\leq 0
\cr
  \frac{e^{-s}\left(1+\beta\right)-1}{\beta}&s\geq 0 }\right..
\end{eqnarray}
Integration over $\mu_{0}$ in Equation (9) yields
\begin{eqnarray}
\mathcal{P} \left(t,p\right)&=&-\frac{3|1-t|}{32p^{6}t}\left[1+
\left(10+8p^{2}+4p^{4}\right)t+t^{2}\right]\nonumber\\
&+&\frac{3\left(1+t\right)}{8p^{5}}\left[\frac{3+3p^{2}+p^{4}}{\sqrt{1
+p^{2}}}-\frac{3+2p^{2}}{2p}\left(2 \sinh^{-1}\left(p\right)
-|\ln\left(t\right)|\right)\right],
\end{eqnarray}
where
\begin{eqnarray}
|\ln\left(t\right)|\leq2 \sinh^{-1}\left(p\right).
\end{eqnarray}

Compton scattering results in a frequency shift, $x\rightarrow x'$, where $x$ 
is the dimensionless frequency $x=h\nu/kT$, and $T$ is the CMB
temperature. The corresponding change of the photon occupation number through 
a pathlength along the cluster, $\Delta n(x)$, is obtained by integrations 
over the scattering probability and electron momentum distributions,
\begin{eqnarray}
\Delta n\left(x\right)=\tau\int_{p_{1}}^{p_{2}}\int_{t_{min}}^{t_{max}}
dpdt \frac{N(p)}{N_{tot}}\mathcal{P} \left(t,p\right)\left(\frac{1}{e^{xt}-1}
-\frac{1}{e^{x}-1}\right),
\end{eqnarray}
where $\tau$ is the optical depth of the cluster to Compton scattering, 
and $N_{tot}$ is the total number of electrons in the cluster.
The measured quantity is the change in intensity, 
$\Delta I(x)/i_{0} = x^{3}\Delta n(x)$, where $i_{0}=2(kT)^{3}/(hc)^{2}$. 
We use the latter equation to calculate the additional intensity change 
induced by energetic electrons, $\Delta I_{nt}(x)$, in the electron models 
described in the previous section, with parameter values sampling the 
observationally deduced ranges, as specified in Table 1 for all the four 
electron models.

If Comptonization by NT electrons is not taken into account it could 
affect the observationally deduced value of the crossover frequency, 
$x_{0}$, defined as the frequency at which the {\it purely thermal} 
effect vanishes. The exact value of $x_{0}$ has practical
significance, since observations near this frequency are optimal for
the determination of cluster peculiar velocities (Rephaeli \& Lahav 1991)
from measurements of the kinematic component of the S-Z effect.
We have calculated the modified value of the frequency where now the
{\it sum} of the thermal and NT intensity changes vanishes. For this we
used the analytic expression of Nozawa, Itoh, \& Kohyama(1998)
for the thermal and
kinematic components of the S-Z effect, and computed $x_{0}$ using the
measured values of the temperatures in Coma and A2199.

\section{Results}

The impact of each of the above four NT electron models depends very much 
on the total number of these electrons as determined by radio, NT EUV 
and X-ray emission. The thermal gas parameters are based on X-ray 
measurements. For Coma, recent XMM measurements yield $kT_{e}=8.2 \pm 0.4$ 
keV for the gas temperature (Arnaud \ea 2001), and (since 
quantitative results for the gas density profile from either XMM or 
{\it Chandra} are not yet available) we have taken the ROSAT deduced 
values (Mohr, Mathiesen, \& Evrard 1999), $n_{e0} \simeq (3.12. \pm 0.04) 
\times 10^{-3}$ cm$^{-3}$, $r_{c}=0.366$ Mpc, and $\beta_{n} \simeq 0.705$, 
for the central electron density, core radius, and index in the expression 
for (the commonly used) density profile, 
$n_{e}(r) = n_{e0}(1+r^{2}/r_{c}^{2})^{-3\beta_{n}/2}$, respectively. 
The corresponding values in A2199 are $kT_{e}=4.7$ keV, central
electron density $n_{e0}=7.21\times 10^{-3}\rm{cm}^{-3}$,
$r_{c}=0.196$ Mpc, and $\beta_{n}=0.78$ (Kaastra et al. 1999). These
values are based on $H_0 = 50$ km s$^{-1}$ Mpc$^{-1}$. 

Clearly, most of the intensity change $\Delta I_{nt}(x)$ is due to the 
more numerous low energy electrons. More specifically, a typical CMB 
photon has dimensionless frequency $\bar{x}=2.701$, and the scattered 
photon frequency is on average
\begin{eqnarray}
\bar{\frac{x'}{x}}=1+\frac{4}{3}p^{2}.
\end{eqnarray}
The observationally relevant range of values of $x$ is at most $x \leq 20$, 
and since the (the 68\% likelihood) interval of rms values of $x'$, 
$\Delta x' \simeq x\sqrt{2 p^{2}/3 + 46p^{4}/45}$, is very wide for 
$p>1$, it follows that electrons with momenta much larger than at most 
a few tens are largely irrelevant for boosting photons to frequencies
such that $x \leq 20$. 
(Obviously, {\it all} electrons contribute to the value of the Thomson 
optical depth and the decrement on the Rayleigh-Jeans side.)

We have calculated $\Delta I_{nt}(x)/i_{0}$ and total electron energies 
in the extended power-law model described by equation (4), using the
observationally deduced model parameters in Coma and A2199. We emphasize 
that the basic model is a power law at high energies ($\geq 1$ GeV) with 
the index determined from radio measurements. The distribution is then 
appropriately extended to lower energies (by taking electronic excitations 
losses into account)
resulting in a change in the value of the index at 
low energies. The power law indices and overall normalizations were taken 
from Rephaeli, Gruber \& Blanco (1999) for Coma, and Kaastra \ea (1999) 
for A2199. Based on these parameters, the deduced values of 
$\Delta I_{nt}(x)/i_{0}$ are very small in comparison with the
magnitude of the 
thermal S-Z effect in these clusters as predicted from the measured 
values of the gas temperature and density. In Coma the S-Z effect was 
actually measured by Herbig et al. (1995) using the 
OVRO 5.5 m telescope, and more recently by De Petris \ea (2002) using 
the MITO telescope. In the former paper the temperature change in the 
center of Coma was reported to be $\Delta T = -505 \pm 92$ $\mu$K at 
$32$ GHz ($x \simeq 0.56$), a value which is consistent with that 
predicted based on the X-ray deduced parameters. A somewhat lower 
value was deduced by De Petris \ea (2002). Our calculated values 
of $\Delta I_{nt}(x)/i_{0}$ due to NT electrons are smaller than
0.001\% of the predicted thermal effect in Coma and A2199. Clearly, 
the implied shift in the value of the crossover frequency is also 
negligible. The very small impact of NT electrons in this model is not 
surprising given the relatively low energy content of these electrons,
$\sim 0.8\%$
of the thermal electron energy in Coma and A2199.

Next we calculated the degree of Comptonization in the shock accelerated 
and cooling electrons models (II and III) described in equations (5) and 
(6). We have normalized these models to match the observed EUV emission, 
at a luminosity level of roughly $5\times 10^{43}\ \rm{erg}\ \rm{s}^{-1}$ over the 
band 0.065-0.245 keV in both Coma and A2199. In the shock accelerated 
electron model the calculated levels of $\Delta I_{nt}(x)$ are $<0.5\%$ 
and $\leq 3\%$ of the corresponding intensity change due to thermal 
electrons, and the shift in the value of the crossover frequency is 
also quite small. Calculated values of the energy in NT electrons in 
this model are $\sim$3\% and 13\%-16\% (for $p_{c}=0.3-1.0$) of the 
energy in thermal electrons in Coma and A2199, respectively. In the 
cooling electron model, the degree of NT Comptonization is negligible, 
less than $0.003\%$ of the magnitude of the thermal S-Z effect in both 
Coma and A2199. This is mainly due to the small number of NT electrons 
in this model, $\sim$ 0.1\% of that in the shock accelerated electron 
model. The NT electron energy constitutes 0.9\% and 4.2\% of 
the electron thermal energy in the two clusters. Our calculated 
quantities are listed in Table 2. The first column is the cluster name 
(or Abell number), the second is the model number, third is the ratio 
of (total) energy in NT to energy in thermal electrons. In the 
fourth column we list the net energy deposition rate (in keV/Gyr) by 
NT electrons, and in the
fifth the value of the crossover frequency; 
values of the error in the peculiar velocity and the ratio 
$\Delta I_{nt}(x)/ \Delta I$ are listed in last two columns.

\begin{table*}[h]
\small
\begin{center}
\begin{tabular}{llllllllll}
\tableline
\tableline
Cluster & Model & $E_{nt}/E_{th}$ & $\quad\rm{dE/dt}$ & $x_{nt}$ & 
$\quad\Delta v>$ & $\Delta I_{nt}/ \Delta I$ \cr
&&\%&(\rm{keV/Gyr})& &(\rm{km/sec})&\qquad\%\cr
\tableline
Coma &&&&&&\cr
     &Thermal& & &3.9000&&&\cr
     &I&0.79&$-0.06$&3.9000&-&$1.4\times 10^{-4}$\cr
     &II&3.18&$10.80$&3.9020&-&0.41\cr
     &III&0.90&$-0.04$&3.9000&-&$4.7\times 10^{-4}$\cr
     &IV&21.59&$296.86$&3.9216&-110&6.8\cr
&&&&&&\cr
A2199&      &    &    &&&\cr
     &Thermal& & &3.8700&&&\cr
     &I&0.74&$-0.15$&3.8701&-&$1.2\times 10^{-4}$\cr
     &II&15.87&$58.87$&3.8803&-&2.97\cr
     &III&4.23&$-0.06$&3.8702&-&$2.5\times 10^{-3}$\cr
     &IV&64.17&$626.39$&3.9385&-210&34.5\cr
\tableline
\end{tabular}
\end{center}
\caption{The impact of the four energetic electron models (see the text for 
definitions of the listed quantities).}
\end{table*}

The much higher NT electron energy content in the single power law (model 
IV) suggested by Sarazin \& Kempner (2000), $\sim 20\% - 22\%$ (for $q=2.92 - 
3.68$) in Coma, and $\sim 64\% - 184\%$ (for $q=2.2 - 3.33$) in A2199, 
result in substantial degrees of additional Comptonization. Using values 
of the parameters as deduced by Sarazin \& Kempner (2000), listed in 
Table 1, we calculate the change of intensity, $\Delta I_{nt}(x)/i_{0}$, 
shown in Figure 1, and listed in Table 2. It is clear from this figure 
that $\Delta I_{nt}(x)$ can reach an appreciable fraction of $\Delta I(x)$. 
For example, near the peak of the Comptonized Planckian, at $x \sim 6.5$, 
$\Delta I_{nt} / \Delta I$ is $\leq 13\%$ and $\leq 35\%$ for Coma and A2199, 
respectively. The implied shift in the value of the crossover frequency in 
this model, $\sim 2$ GHz in Coma, and $\geq 4$ GHz in A2199, would also 
be observationally relevant. Such a systematic shift would introduce an 
error in the deduced value of the peculiar velocity. The implied 
error (which does not depend on the velocity, if the very small quadratic 
dependence of $\Delta I(x)$ on the velocity is ignored) amounts to 
$\sim 110$ km/s for Coma ($q= 2.92 - 3.68$), and $\sim 210$ km/s 
($q=3.33$) and $320$ km/s ($q=2.2$) for A2199 (values lower than 
100 km/s are not shown in Table 2).

\begin{figure}[h]
\plottwo{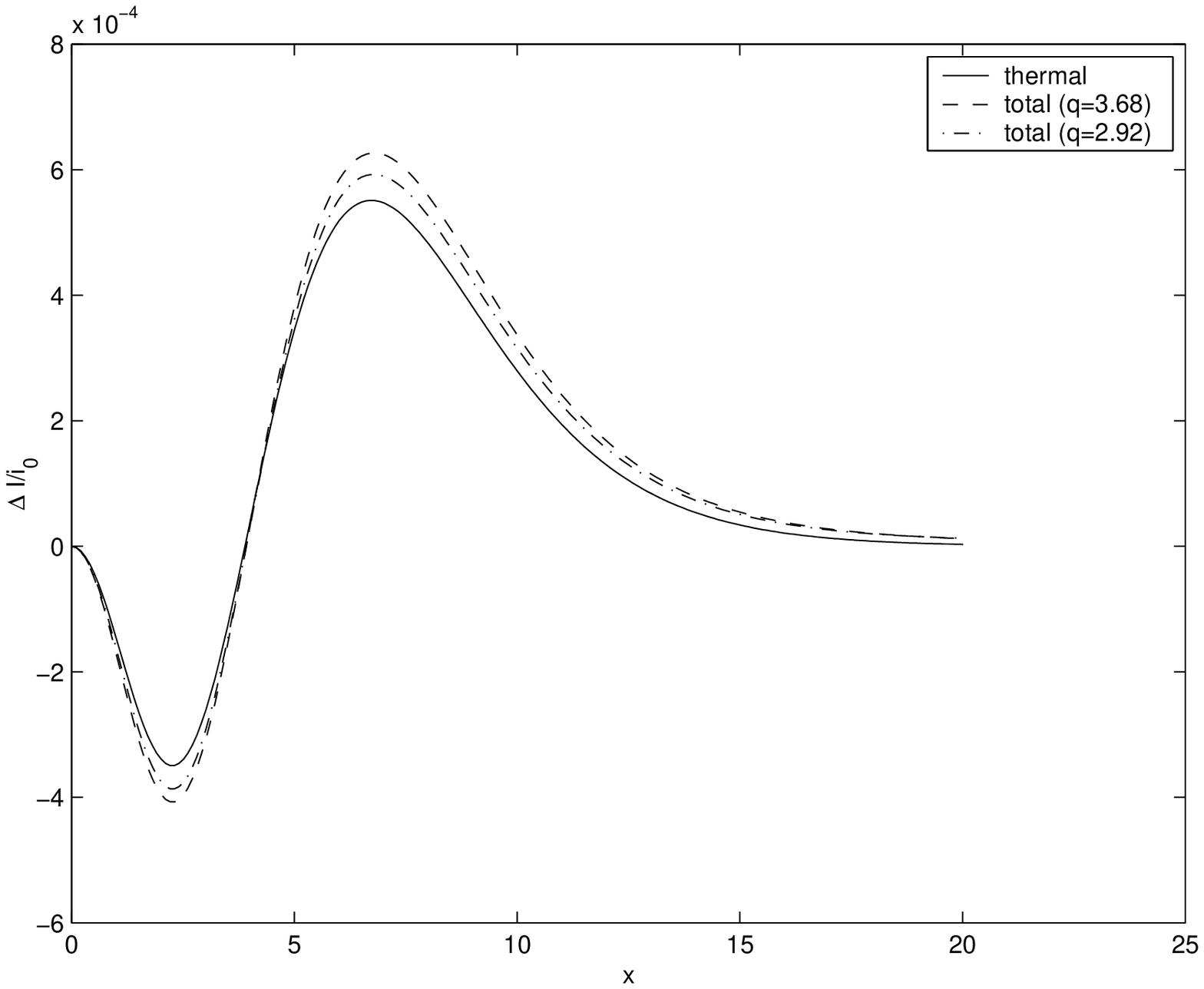}{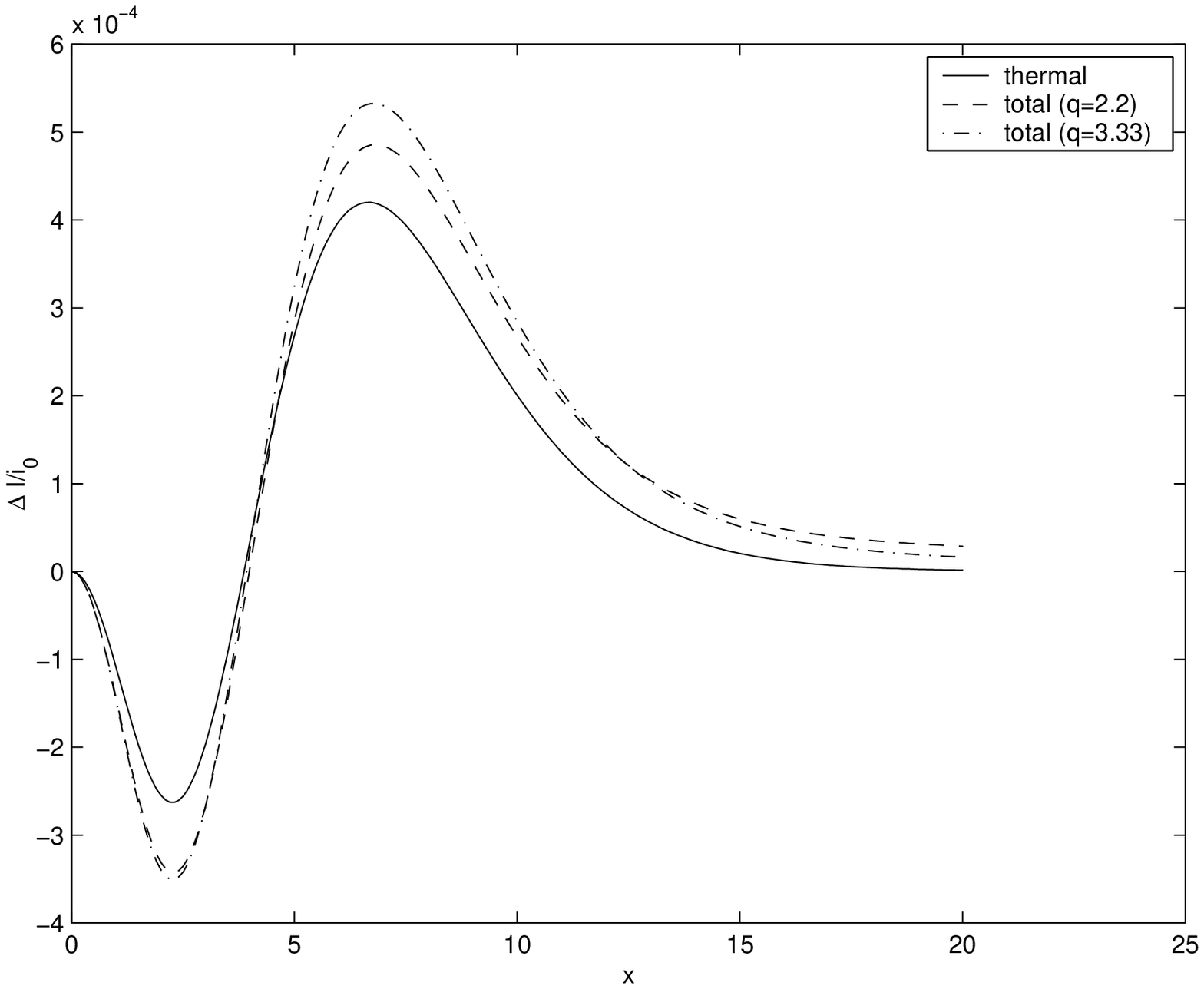}
\caption{The spectrum of the additional Comptonization due to energetic 
electrons in the models proposed by Sarazin \& Kempner (2000) for the 
Coma (left panel) and A2199 (right panel).}
\end{figure}

Finally, we have calculated the heating rate of the hot gas by the NT 
electrons through Coulomb interactions (Rephaeli 1979, Rephaeli \& Silk 
1995) in the four models considered in this paper. Energy deposition rates 
into the gas were estimated by taking a mean gas density in the central 
1 Mpc region of Coma and A2199. Subtraction of the observed cooling rate 
due to thermal bremsstrahlung (X-ray) emission then yields the net rate of 
temperature change. Values of the latter quantity (in keV/Gyr) are listed 
in Table 2. 

\section{Discussion}

We have briefly discussed models for NT energetic electrons and 
their predicted effect on the CMB based on parameters that were 
deduced from radio, EUV, and X-ray measurements of Coma and A2199. 
Our immediate objective has been the exact calculation of the additional
degree of Comptonization due to these electrons, and its possibly relevant 
observational consequences. Only electron models that have direct
observational motivation were considered. Since all these models have
basically a power-law form, it is clear that their impact on the CMB is 
largely due to the more abundant lower energy electrons (except for the 
cooling electrons model). However, the interpretation of the EUV and X-ray 
emissions as NT bremsstrahlung and Compton scattering are not secure, so 
there is considerable uncertainty in quantifying electron densities. 
Moreover, since low energy electrons lose energy mostly by non-radiative 
coupling to the thermal gas, the uncertainty in the estimation of their 
density is particularly substantial. Only the presence of \rel electrons is 
well established from measurements of extended regions of radio emission 
in many clusters (including Coma, but not A2199). Clearly, radio 
measurements yield the electron density only if we have an independent 
estimate of the mean, volume-averaged value of the magnetic field in the 
emitting region. Estimates of the mean IC field are few and uncertain 
(Newman, Newman \& Rephaeli 2002).

Due to the lack of detailed information on IC energetic electron 
populations, relevant theoretical considerations are of particular 
interest. First, since it is virtually always the case that particle 
acceleration mechanisms tap only a small fraction of the energy at the 
source, energy density in NT electrons is likely to be only a small 
fraction of the gas (which is a significant mass component of clusters) 
thermal energy density. Realistically, therefore, the thermal energy 
density in clusters is expected to set an absolute upper limit to the 
energy density in NT electrons, a limit which is very unlikely to be 
reached. Second, the energetic electron population is linked to the 
gas mainly by Coulomb coupling that results in energy transfer from 
energetic electrons (and protons) to the gas. This coupling and the 
resulting heating set stringent constraints on the attainment of 
steady state in general, and the density of low energy NT electron 
models in particular. 

We have calculated the ratio of total energy in NT electrons to that 
in thermal electrons in the four energetic electron models 
considered in this paper; for the first three models the calculated 
values (in Table 2) are low. For the power law model of Sarazin \& Kempner 
(2000), the corresponding values are very high ($\sim 22\%$ in Coma, and 
$\geq 64\%$ in A2199) and quite unrealistic. Such large energy contents 
would also imply a high rate of energy transfer to the gas, and heating at 
a rate higher than the rate of cooling by emission of thermal radiation. 
This without even considering the additional heating by IC energetic 
protons (whose energy density in the Galaxy is higher than that of \rel 
electrons). 

In fact, the single power law model of Sarazin \& Kempner (2000) is not 
only unappealing from an energetic point of view, but is also questionable 
on a more basic physical ground: As we have stated in Section 2 -- based 
on the original work of Rephaeli (1979) -- correct extrapolation of the 
\rel electron energy spectrum to low ($<100$ MeV) energies must account 
for electronic Coulomb excitations, the dominant energy loss mechanism 
at such energies for typical gas densities in the central regions of 
clusters. The very weak energy dependence of this mechanism (in contrast 
with the quadratic dependence of the Compton-synchrotron energy loss rate) 
flattens the spectrum at low energies, resulting in a much smaller number 
of electrons than would have been predicted otherwise. We conclude that 
this model is non-viable; therefore, the high degree of implied 
Comptonization and appreciable shift in the value of $x_0$ that 
are predicted in this model are at best very high upper limits to 
the impact of NT electrons on measurements of the S-Z effect.

Our basic result in this paper is that exact calculation of the impact 
of realistically normalized models of NT electrons in Coma and A2199 yields 
levels of the degree of Comptonization by such electrons that are only a 
negligible fraction of the corresponding S-Z effect due to the hot IC 
gas. It is clear from our discussion that this result is generally valid 
since the energy density in NT electrons is not likely to be a significant 
fraction of the thermal energy density. In clusters with a significant NT 
electron population the added Comptonization due to these electrons 
would clearly constitute a source of confusion in the analysis of S-Z 
measurements. Since there are little or no differences in the spectral 
CMB signatures of thermal and NT electrons, high spatial resolution X-ray, 
S-Z and radio measurements of these clusters would be needed in order 
to minimize this confusion when (as expected) the spatial profiles of the 
two electron populations are found to be detectably different.

\acknowledgments
Useful comments made by the referee on an earlier version of the paper
are gratefully acknowledged. This research has been supported by the
Israeli Science Foundation grant 729$\backslash$00-02 at Tel Aviv University.


\begin{thebibliography}{99}
\bibitem{1} Arnaud, M.~et al.\ 2001, \aap, 365, L67
\bibitem{2} Baring, M.~G., Ellison, D.~C., Reynolds, S.~P., Grenier, 
I.~A., \& Goret, P.\ 1999, \apj, 513, 311 
\bibitem{3} Birkinshaw, M.\ 1999, \physrep, 310, 97
\bibitem{4} Blasi, P.~\& Colafrancesco, S.\ 1999, Astroparticle Physics, 
12, 169 
\bibitem{5} Blasi, P., Olinto, A.~V., \& Stebbins, A.\ 2000, \apjl, 535, L71 
\bibitem{6} Bowyer, S., Bergh{\" o}fer, T.~W., \& Korpela, E.~J.\ 1999, 
\apj, 526, 592 
\bibitem{61} Carlstrom, J.~E.~\& et al.\ 2000, IAU Symposium, 201,
  E48
\bibitem{7} Carlstrom, J.E. \ea 2001, astro-ph/0103480
\bibitem{8} Challinor, A.~\& Lasenby, A.\ 1998, \apj, 499, 1 
\bibitem{9} Chandrasekhar, S.\ 1950, Oxford, Clarendon Press, 1950
\bibitem{10} De Petris, M. \ea 2002, astro-ph/0203303
\bibitem{11} Ensslin, T.~A.~\& Kaiser, C.~R.\ 2000, \aap, 360, 417 
\bibitem{12} Fusco-Femiano, R., dal Fiume, D., Feretti, L., Giovannini, G., 
Grandi, P., Matt, G., Molendi, S., \& Santangelo, A.\ 1999, \apjl, 513, L21 
\bibitem{13} Giovannini, G., Tordi, M., \& Feretti, L.\ 1999, New 
Astronomy, 4, 141 
\bibitem{14} Giovannini, G.~\& Feretti, L.\ 2000, New Astronomy, 5, 335 
\bibitem{15} Gruber D.E., \& Rephaeli Y. 2002, \apj, in press 
(astro-ph/0110512) 
\bibitem{16} Herbig, T., Lawrence, C.~R., Readhead, A.~C.~S., \& Gulkis, 
S.\ 1995, \apjl, 449, L5 
\bibitem{17} Itoh N., Kawana Y., Nozawa S., Kohyama Y.,
  astro-ph/0005390
\bibitem{151} Jones, M., et al. 1993, Naure, 365, 320
\bibitem{18} Kaastra, J.~S., Lieu, R., Mittaz, J.~P.~D., Bleeker, J.~A.~M., 
Mewe, R., Colafrancesco, S., \& Lockman, F.~J.\ 1999, \apjl, 519, L119 
\bibitem{19} McKinnon, M.~M., Owen, F.~N., \& Eilek, J.~A.\ 1991, \aj, 101, 
2026 
\bibitem{20} Mohr, J.~J., Mathiesen, B., \& Evrard, A.~E.\ 1999, \apj, 517, 
627 
\bibitem{21} Molendi, S., De Grandi, S., \& Fusco-Femiano, R.\ 2000, \apjl, 
534, L43 
\bibitem{211} Molnar, S.~M.~\& Birkinshaw, M.\ 1999, \apj, 523, 78. 
\bibitem{22} Newman, W.I., Newman, A.L., \& Rephaeli, Y. 2002, \apj, in press
\bibitem{23} Nozawa, S., Itoh, N., \& Kohyama, Y.\ 1998, \apj, 508, 17 
\bibitem{24} Rephaeli, Y.\ 1977, \apj, 212, 608 
\bibitem{25} Rephaeli, Y.\ 1979, \apj, 227, 364 
\bibitem{26} Rephaeli, Y.~\& Lahav, O.\ 1991, \apj, 372, 21 
\bibitem{27} Rephaeli, Y.\ 1995a, \araa, 33, 541 
\bibitem{28} Rephaeli, Y.\ 1995b, \apj, 445, 33 
\bibitem{29} Rephaeli, Y.~\& Silk, J.\ 1995, \apj, 442, 91 
\bibitem{30} Rephaeli, Y.~\& Yankovitch, D.\ 1997, \apjl, 481, L55 
\bibitem{31} Rephaeli, Y., Gruber, D., \& Blanco, P.\ 1999, \apjl, 511, L21 
\bibitem{32} Rephaeli Y. 2001, {\it Astrophysical Sources of High Energy 
Particles \& Radiation}, Kluwer, Dordrecht (astro-ph/0105265) 
\bibitem{33} Sarazin, C.~L.~\& Lieu, R.\ 1998, \apjl, 494, L177
\bibitem{331} Sarazin, C.~L.\ 1999, \apj, 520, 529.  
\bibitem{34} Sarazin, C.~L.~\& Kempner, J.~C.\ 2000, \apj, 533, 73 
\bibitem{35} Sazonov, S.~Y.~\& Sunyaev, R.~A.\ 1998, \apj, 508, 1
\bibitem{36} Shimon M., Rephaeli Y.\ 2002, in preparation 
\bibitem{37} Sunyaev, R.~A.~\& Zeldovich Y.~B.\ 1972, \casp, 4, 173
\bibitem{38} Sunyaev R.~A.~\& Zel'dovich Y.~B.\ 1980, \mnras, 190, 413
\bibitem{381} Udomprasert, P.~S., Mason, B.~S., \& Readhead, A.~C.~S.\
  2000, American Astronomical Society Meeting, 197, 110401 
\bibitem{39} Wright, E.~L.\ 1979, \apj, 232, 348 
\bibitem{40} Yamada, M., Sugiyama, N., \& Silk, J.\ 1999, \apj, 522, 66 
\bibitem{41} Zeldovich, Y.~B.~\& Syunaev, R.~A.\ 1969, \apss, 4, 301 
\end{thebibliography}
\end{document}